\documentclass{IEEEtran}
\usepackage{amssymb,amsmath} 
\usepackage{graphicx}
\usepackage{url}
\usepackage[colorlinks=true, pdfstartview=FitV, linkcolor=blue, citecolor=blue, urlcolor=blue]{hyperref}
\usepackage{hyperref}

\def\={\equiv} 
\def\div{\nabla\cdot } 
\def\pl{\partial}

\newcommand{\rt}{(\3r,t)}

\newcommand{\ox}{(\3x)}

\newcommand{\bib}{\bibitem}

\newcommand{\ci}{\cite}
\newcommand{\lab}{\label}

\newcommand{\eq}{\eqref}

\newcommand{\harr}[1]{\smash{\mathop{\hbox to .5in{\ \rightarrowfill\ }}
      \limits^{#1}}}

\newcommand{\3}[1]{{\boldsymbol #1}}

\newcommand{\bh}[1]{{\boldsymbol{\hat #1}}}

\newcommand{\9}[1]{^{\,\scriptscriptstyle#1}}

\def\e{\varepsilon}

\def\m{\mu} 

\def\o{\omega} 
\def\p{\pi}

\newcommand{\db}{{\,{\rm d}\kern-1.6ex-}}
\newcommand{\dir}{{\pl\kern-1.2ex {/}}}

\newcommand{\ie}{{\it i.e., }}

\def\iff{\ \Leftrightarrow\ }

\newcommand{\lra}{\leftrightarrow}

\newcommand{\plra}{\pl^{\kern-1.25ex^\lra}}
\newcommand{\qq}{\quad} 
\newcommand{\qqq}{\qquad}

\newcommand{\sr}{\sqrt}

\def\XXint#1#2#3{{\setbox0=\hbox{$#1{#2#3}{\int}$}
     \vcenter{\hbox{$#2#3$}}\kern-.5\wd0}}

\def\HB{\hfill\break}

\def\ED{\end{document}}

\def\bib#1{\bibitem{#1}}

\begin{document}

\title{The Reactive Energy of Transient EM Fields}

\author{{\Large Gerald Kaiser\,*\thanks{*Supported by AFOSR Grant \#FA9550-08-1-0144.}}\\
\href{http://wavelets.com}{Center for Signals and Waves}\\ Austin, TX\\
kaiser@wavelets.com
}

\maketitle

\pagestyle{empty}\thispagestyle{empty}


\section{The reactive energy density $\5R\rt$}

We give a physically compelling definition of the \sl instantaneous reactive energy density \rm associated with an arbitrary time-domain electromagnetic field in vacuum \ci{K11}. In \sl Heaviside-Lorentz units, \rm where $\e_0=\m_0=1$, it is given in terms of the energy density $U\rt$ and the Poynting vector $\3S\rt$  by
\begin{align}\lab{R}
\5R\rt=\sr{U\rt^2-\3S\rt^2}.
\end{align}
This is a field-theoretic version of the \sl rest energy of a relativistic point particle \rm  with total energy $E$ and momentum $\3p$, 
\begin{align*}
E_0=\sr{E^2-c^2\3p^2}.
\end{align*}
We may interpret \eq{R} as follows: at space-time points $\rt$ where $|\3S|<U$, the energy flow is insufficient to carry away \sl all \rm of the energy in the form of radiation. The (momentarily) abandoned `rest' energy is  reactive. 

In terms of the electric and magnetic fields $(\3E,\3B)$, we have
\begin{align*}
U=\tfrac12(\3E^2+\3B^2),\qqq \3S=\3E\times\3B
\end{align*}
and $\5R$ reduces to the simple expression
\begin{align}\lab{REB}
\5R=\sr{\tfrac14(\3E^2-\3B^2)^2+(\3E\cdot\3B)^2}\ge 0.
\end{align}
This shows that at each space-time point $\rt$ we have
\begin{align}\lab{R1}
\5R=0\iff \3E^2-\3B^2=0\ \ \hbox{and}\ \ \3E\cdot\3B=0,
\end{align}
which are precisely the conditions for a \sl pure radiation field. \rm For a \sl generic \rm EM field, \sl $\5R$ is strictly positive almost everywhere\footnote{Here \sl almost everywhere \rm means that $\5R$ can vanish only on lower-dimensional hypersurfaces of space-time. If $\3E^2-\3B^2$ and $\3E\cdot\3B$ are \sl independent, \rm  \eq{R1} implies that $\5R=0$ on a 2D space-time surface whose time slices are, in general, time-dependent curves in space. For the standing plane wave in Example 2 below, $\3E\cdot\3B\equiv0$, so \eq{R1} reduces to one condition and $\5R$ vanishes on the traveling planes \eq{R0}, which form 3D hypersurfaces in space-time whose time slices are snapshots of the planes at a given time $t$.
}
in space-time and approaches zero, as it must, only in the far zone. \rm Fields for which $\5R$ vanishes identically, called \sl null fields, \rm consist of pure radiation. The simplest null fields are traveling plane waves. An interesting example of null fields with sources, resembling a spinning black hole in general relativity,  was constructed in \ci{K11a}. It was this example that inspired the general study of reactive energy density in \ci{K11}.

Just as the rest energy $E_0$ defines the \sl mass \rm $m$ of the point particle  by $E_0=mc^2$, so does $\5R$ define the \sl electromagnetic inertia density \rm $\5I$ by
\begin{align*}
\5R\rt=\5I\rt c^2.
\end{align*}
Whereas $m$ and $E_0$ measure \sl impedance to acceleration, \rm $\5I$ and $\5R$ measure \sl impedance to radiation. \rm Like $E_0$, $\5R$ is \sl Lorentz invariant, \rm \ie it has identical values in all uniformly moving (inertial) coordinate frames. For \sl narrowband \rm fields, the time average of $\5R$ is expected to reduce to the known, stationary reactive energy density. Thus $\5R$ is a transient or `ultra-wideband' version of the latter, local in \sl time \rm as well as space. 

We compute $\5R\rt$ explicitly for two fields representing the extremes of space-time localization:
\begin{enumerate}
\item A general time-dependent electric dipole field. This is \sl local \rm in space-time.
\item A standing plane wave obtained by adding two plane waves of frequency $\o>0$ traveling along $\pm\bh z$. This is localized at two points in the 4D frequency-wavenumber domain, hence \sl highly nonlocal \rm in space-time.
\end{enumerate}
In Example 1, we find that the reactive energy oscillates around the dipole, as shown in Figure 1, and decays to zero in the far zone as expected. \HB
\begin{figure}[h]
\begin{center}
\includegraphics[width=1.2 in]{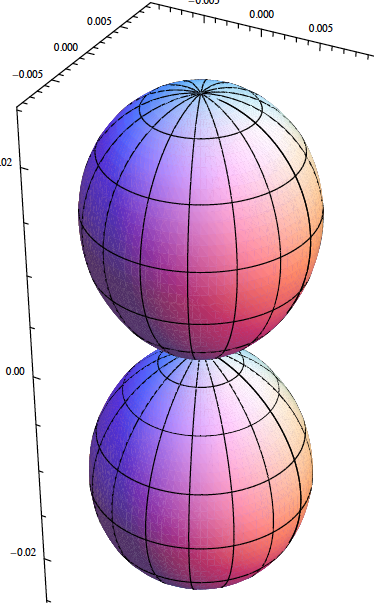}\ \ 
\includegraphics[width=1.6 in]{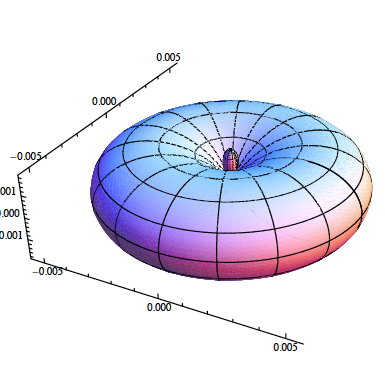}
\caption{\small For a frequency-modulated Gaussian electric dipole, the near-field pattern of the reactive energy density oscillates between the two forms shown above; see \ci{K11} for details.}
\end{center}
\end{figure}

In Example 2, we have 
\begin{align}\lab{US}
\begin{aligned}
U&=E^2[\cos^2(kz-\o t)+\cos^2(kz+\o t)]\\
\3S&=\bh zE^2[\cos^2(kz-\o t)-\cos^2(kz+\o t)],
\end{aligned}
\end{align}
where $k=\o/c$ and $E$ is the amplitude of the electric fields of the traveling plane waves. 
This gives
\begin{align}\lab{RSW}
\5R=2E^2|\cos(kz-\o t)\cos(kz+\o t)|.
\end{align}
Thus $\5R$ vanishes on the \sl traveling nodal planes \rm $z=z_\ell\9\pm\0t$, where
\begin{align}\lab{R0}
z_\ell\9\pm\0t=\frac{(2\ell+1)\p}{2k}\pm ct,\qq \ell=0, \pm 1, \pm 2,\cdots ,
\end{align}
and $\5R>0$ elsewhere.  This is shown in Figure 2.
\begin{figure}[h]
\begin{center}
\includegraphics[width=3.3 in]{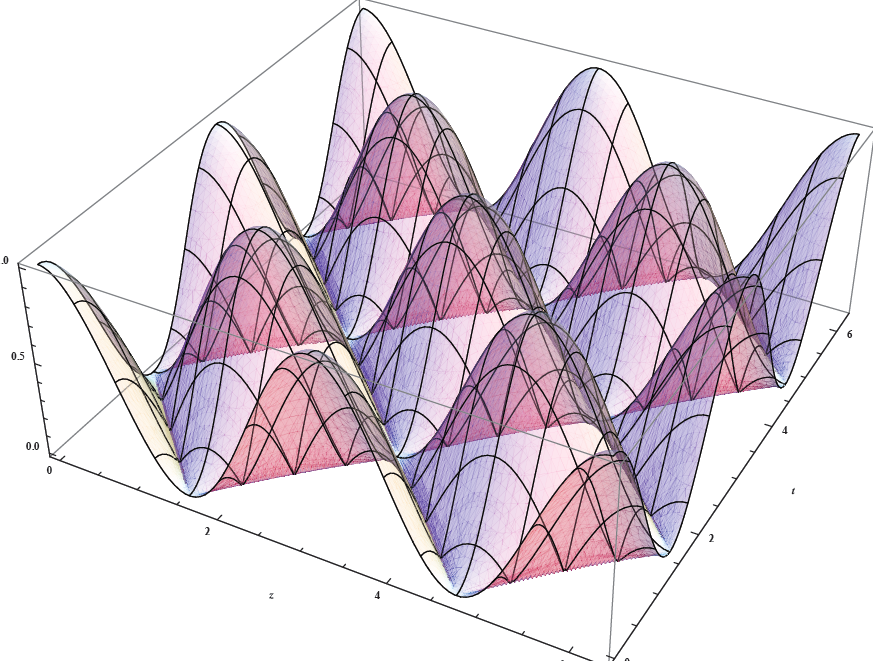}
\caption{\small The reactive energy density $\5R(z,t)$ of the standing plane wave in Example 2, showing the moving nodes \eq{R0} where $\5R=0.$}
\end{center}
\end{figure}

The two plane waves traveling along $\pm\bh z$ are \sl null, \rm \ie their reactive energy densities  vanish. Hence the reactive energy of their sum, the standing wave, is due entirely to the \sl interference \rm between the two traveling waves. That is, the invariants in \eq{R1} consist only of the cross-terms.  Furthermore, since every \sl globally sourceless \rm field is a Fourier superposition of null plane waves with $\o=ck$, it follows that \sl the reactive energy of every globally sourceless EM field is due entirely to self-interference. \rm This gives a \sl partial \rm intuitve explanation of EM rest energy, as seen most clearly in the standing wave example. However, the rest energy of fields \sl with \rm sources need \sl not \rm be entirely due to self-interference since their Fourier synthesis also requires plane waves with $\o\ne ck$, which are not null. (Such plane waves represent `virtual photons,' which have positive mass.)

\section{The energy flow velocity $\3v\rt$}

The correspondence between the rest energy $E_0$ of a relativistic point particle and the reactive energy density $\5R\rt$ of an EM field in vacuum can be extended to include the \sl velocity \rm of the point particle, 
\begin{align}\lab{v}
\3v=\frac{c^2\3p}E,
\end{align}
whose field-theoretic version is
\begin{align}\lab{vxt}
\3v\rt=\frac{c\3S\rt}{U\rt}.
\end{align}
Poynting's theorem $\pl_t U+c\div\3S=-\3J\cdot\3E$ then becomes
\begin{align}\lab{Poynt1}
\pl_t U+\div(\3v U)=-\3J\cdot\3E,
\end{align}
which shows that \sl $U$ behaves like the density of a compressible fluid with source $-\3J\cdot\3E$, flowing at velocity $\3v\rt$. \rm Note that
\begin{align}\lab{vc}
v\rt\=|\3v\rt|=c\iff \5R\rt=0.
\end{align}
Thus, while the \sl field \rm $(\3E,\3B)$ propagates at $c$, its \sl energy \rm generally flows at $v<c$ almost everywhere. \rm 

For the standing plane wave of Example 2, \eq{US} shows that $|\3v|\le c$ as expected, and
\begin{align*}
\3v=\30&\iff \cos^2(kz-\o t)=\cos^2(kz+\o t)\\
&\iff kz+\o t=\pm(kz-\o t)+n\p.
\end{align*}
Hence $\3v$ has \sl fixed nodes in both space and time: \rm 
\begin{align}\lab{nodes2}
\3v=\30\iff z=\frac{n\p}{2k}\=z_n\ \ \hbox{or}\ \ \ t=\frac{n\p}{2\o}\= t_n
\end{align}
where $n$ is any integer. Since $\3v$ changes sign at $z_n$ and $t_n$, the energy is \sl totally reflected \rm at these nodes.

\it The energy oscillates back and forth between the nodal planes $z=z_n$, and $v(z,t)\=\bh z\cdot\3v$ oscillates between $\pm c$ at any $z$. \rm

The conflict between the moving nodes \eq{R0}, where $v=\pm c$, and the stationary nodes \eq{nodes2}, where $v=0$, is resolved by noting that $v(z,t)$ is \sl undefined \rm when $U=0$ and $\3S=\30$, so
\begin{align*}
&\cos(kz-\o t)=\cos(kz+\o t)=0.
\end{align*}
This gives $\cos kz\cos\o t=0$ and $\sin kz\sin\o t=0$, hence
\begin{align*}
z=z_n\ \hbox{and}\  t=t_m\ \hbox{with $m+n$ odd}.
\end{align*}
These planes are the \sl intersections \rm of the traveling and stationary nodes. Intuitively, the reason why $v(z,t)$ is undefined at these events is that \sl perfect reflection \rm  there requires it to change \sl instantaneously \rm between the values $\pm c$. At all other values of $z$, $v(z,t)$ still oscillates between $\pm c$ but does so in a continuous manner; see Figure 6 in \ci{K11}.

\section{A historical note}

The fact that the energy of an EM field in vacuum generally flows at speeds less than $c$ was noted almost a century ago by Bateman \ci[page 6]{B15}. To the best of my knowledge, this important insight has remained undeveloped and largely unappreciated.\footnote{For a \sl time-harmonic \rm field of frequency $\o$, the \sl energy transport velocity \rm is commonly defined as $\3v_\o\ox=c\3S_\o\ox/U_\o\ox$, where $\3S_\o\ox$ and $U_\o\ox$ are the \sl time averages \rm of $\3S\rt$ and $U\rt$ over one period $2\p/\o$. In general, $|\3v_\o\ox|<c$ almost everywhere. However, time-averaging is lossy and the ratio of two averages is not the average of the ratio. Hence $\3v_\o$ is not a time average of the \sl exact, instantaneous \rm energy flow velocity  $\3v\rt$. I thank Professor Andrea Alu for pointing this out.
}
I believe this phenomenon, and its relation to reactive energy as detailed in \ci{K11}, are fundamental features of electromagnetic fields which ought to be studied both theoretically and  experimentally.

%

\end{document}